\documentclass[
               showpacs,            
               preprintnumbers,     
               aps,                 
               prd,                 
               eqsecnum,            
               superscriptaddress,  
               nofootinbib         
               ,tightenlines         
               ]{revtex4}

\usepackage{graphicx}
\usepackage{amssymb,latexsym}

\newcommand{\beq}{\begin{equation}}
\newcommand{\beqa}{\begin{eqnarray}}
\newcommand{\eeq}{\end{equation}}
\newcommand{\eeqa}{\end{eqnarray}}
\newcommand{\p}{\phi}

\newcommand{\siml}{\lesssim}

\def\PLBB#1#2#3{Phys. Lett. B {\bf #1}, #2 (20#3)}

\begin{document}
\preprint{astro-ph/0510598
}

\title{$w$ and $w'$ of Scalar Field Models of Dark Energy
}

\author{Takeshi Chiba}%
\affiliation{
Department of Physics, College of Humanities and Sciences, \\
Nihon University, 
Tokyo 156-8550, Japan}
\affiliation{%
Division of Theoretical Astronomy, National Astronomical Observatory of 
Japan, Tokyo 181-8588, Japan }

\date{\today}


\begin{abstract}
Important observables to reveal the nature of dark energy are 
the equation of state $w$ and its time derivative in units of 
the Hubble time $w'$. Recently, it was shown that the simplest  
scalar field models of dark energy (quintessence) occupy 
rather narrow regions in the $w-w'$ plane.  
We extend the $w-w'$ plane to $w<-1$ and derive bounds on $w'$ 
as a function of $w$ for tracker phantom dark energy. 
We also derive bounds on tracker k-essence. 
\end{abstract}

\maketitle

\section{Introduction}

The equation of state $w=p/\rho$ of dark energy is a key observable to 
reveal the nature of dark energy which accelerates the Universe. 
$w=-1$ for the cosmological constant and $w$ is in general a function of 
time for scalar field models of dark energy (quintessence). $w$ and its time 
derivative in units of the Hubble time, $w'=dw/d\ln a$, are currently 
constrained from the distance measurements of SNIa (assuming a 
prior on $\Omega_m$) as $w_0=-1.31\pm^{0.22}_{0.28},w'_0=-1.48\pm^{0.90}_{0.81}$ 
(at 95\% confidence level) \cite{riess}. 
Important question is how much precisely 
we should determine the equation of state observationally. 

In this respect, Caldwell and Linder have recently attempted an 
observation-oriented phase space analysis of quintessence \cite{cl}. 
Namely, instead of the scalar field and its time derivative, they 
numerically studied the dynamics in the $w-w'$ plane and found that 
``phase space'' of quintessence in the $w-w'$ plane is narrow 
and that a desired measurement resolution should be  
$\sigma(w')\siml (1+w)$. More recently, Scherrer derived analytically 
a tighter lower bound on $w'$ \cite{sch}. 
In this paper, after reviewing the limits (Sec.2) and slightly updating   
the result in \cite{sch},  
we extend these results to phantom dark energy (Sec.3) 
and k-essence (Sec.4).

\section{Limits of Quintessence Revisited}

Firstly, we review the limit of tracker quintessence \cite{sch} to introduce 
the notation. Then we obtain a lower bound on $w'$ for tracker quintessence models.

\subsection{Generic Bound}

We consider a flat universe consisting of (nonrelativistic) matter and scalar 
field dark energy $\phi$ (quintessence). 
The equation of motion of quintessence field $\phi$ is 
\beqa
\ddot\phi+3H\dot\phi+V_{,\phi}(\phi)=0,
\label{eq:eom}
\eeqa
where $V_{,\phi}=\delta V/\delta\phi$. 
The equation of state $w$ is given by 
\beq
w={\dot\phi^2/2-V\over \dot\phi^2/2+V}.
\label{w}
\eeq
Eq.(\ref{w}) suggests that the equation of motion 
Eq.(\ref{eq:eom}) may be rewritten by using $w'=dw/d\ln a$. 
In fact, it is rewritten as \cite{swz}
\beq
\mp{V_{,\phi}\over V}=\sqrt{3\kappa^2(1+w)\over \Omega_{\p}}
\left(1+{x'\over 6}\right),
\label{eomx}
\eeq
where the minus sign corresponds 
to $\dot\phi>0(V_{,\phi}<0)$ and the plus sign 
to the opposite, $\kappa^2=8\pi G$, and $\Omega_{\phi}$ is 
the density parameter of dark energy. $x$ is defined by
\beq
x=\ln\left({1+w\over 1-w}\right),
\eeq
and $x'$ is the derivative of $x$ with respect to $\ln a$ and is 
related with $w'$ as
\beq
x'={2w'\over (1-w)(1+w)}.
\label{x'}
\eeq
Since the left-hand side of Eq.(\ref{eomx}) is positive, $1+x'/6>0$. 
In terms of $w'$ by the use of Eq.(\ref{x'}), we obtain \cite{sch} 
\beq
w'>-3(1-w)(1+w).
\label{quint}
\eeq
This bound applies to a more general class of quintessence field which 
monotonically rolls down the potential. 

\subsection{Tracker Quintessence}

The bound can be tightened for tracker fields which have nearly constant $w$ 
initially and eventually evolve toward $w=-1$. Tracker fields 
have attractor-like solutions in the sense that a very wide range of
initial conditions rapidly converge to a common cosmic evolutionary 
track \cite{swz}. Taking the derivative of Eq.(\ref{eomx}) with respect to 
$\phi$, we obtain \cite{rubano,sch}
\beqa
\Gamma-1={3(w_B-w)(1-\Omega_{\p})\over (1+w)(6+x')}-{(1-w)x'\over 2(1+w)(6+x')}
-{2x''\over (1+w)(6+x')^2},
\label{q-trac}
\eeqa
where $\Gamma=VV_{,\phi\phi}/V_{,\phi}^2$, $w_B$ is the equation of state of background matter 
 and $x''$ is the second derivative of $x$ with respect to $\ln a$. 
Since $w$ is a constant for tracker fields and $\Omega_{\p}$ is 
initially negligible, $w$ is written in terms of $\Gamma$ as
\beq
w={w_B-2(\Gamma-1)\over 2(\Gamma -1)+1}.
\label{eosq}
\eeq
Since $w'\leq 0$ for tracker fields, $x'\leq 0$. However, since $w$ 
asymptotically approaches toward $-1$, $x'$ eventually stops decreasing and 
then increases toward zero. The minimum of $x'$, $x_m'$, gives the minimum of 
$w'$ via Eq.(\ref{x'}). To find $x_m'$, we put $x''=0$ in Eq.(\ref{q-trac}) 
and find that
\beqa
x_m'=-6{w(1-\Omega_{\phi})+2(1+w)(\Gamma -1)\over 
(1-w)+2(1+w)(\Gamma -1)}>-6{2(1+w)(\Gamma -1)\over 
(1-w)+2(1+w)(\Gamma -1)}.
\eeqa
Since $x_m$ is a decreasing 
function of $w(>-1)$, a lower bound is given by 
$w$ of the tracker solution Eq.(\ref{eosq})
\beq
x_m'>{6w\over 1-2w}.
\eeq
{}From Eq.(\ref{x'}), in terms of $w'$, we obtain a lower bound on $w'$
\beqa
w'>{3w\over 1-2w}(1-w)(1+w)\geq -(1-w)(1+w).
\label{q-track2}
\eeqa
The last inequality is the limit derived by Scherrer \cite{sch}. 
In \cite{sch}, $\Gamma >1$ is used to derive the final inequality. However, 
for tracker quintessence, $\Gamma =1-w/2(1+w)(>1)$ and thus a slightly 
stronger bound is obtained. These bounds are shown in Fig.1 and Fig. 2.

\begin{figure}
\includegraphics[width=\hsize]{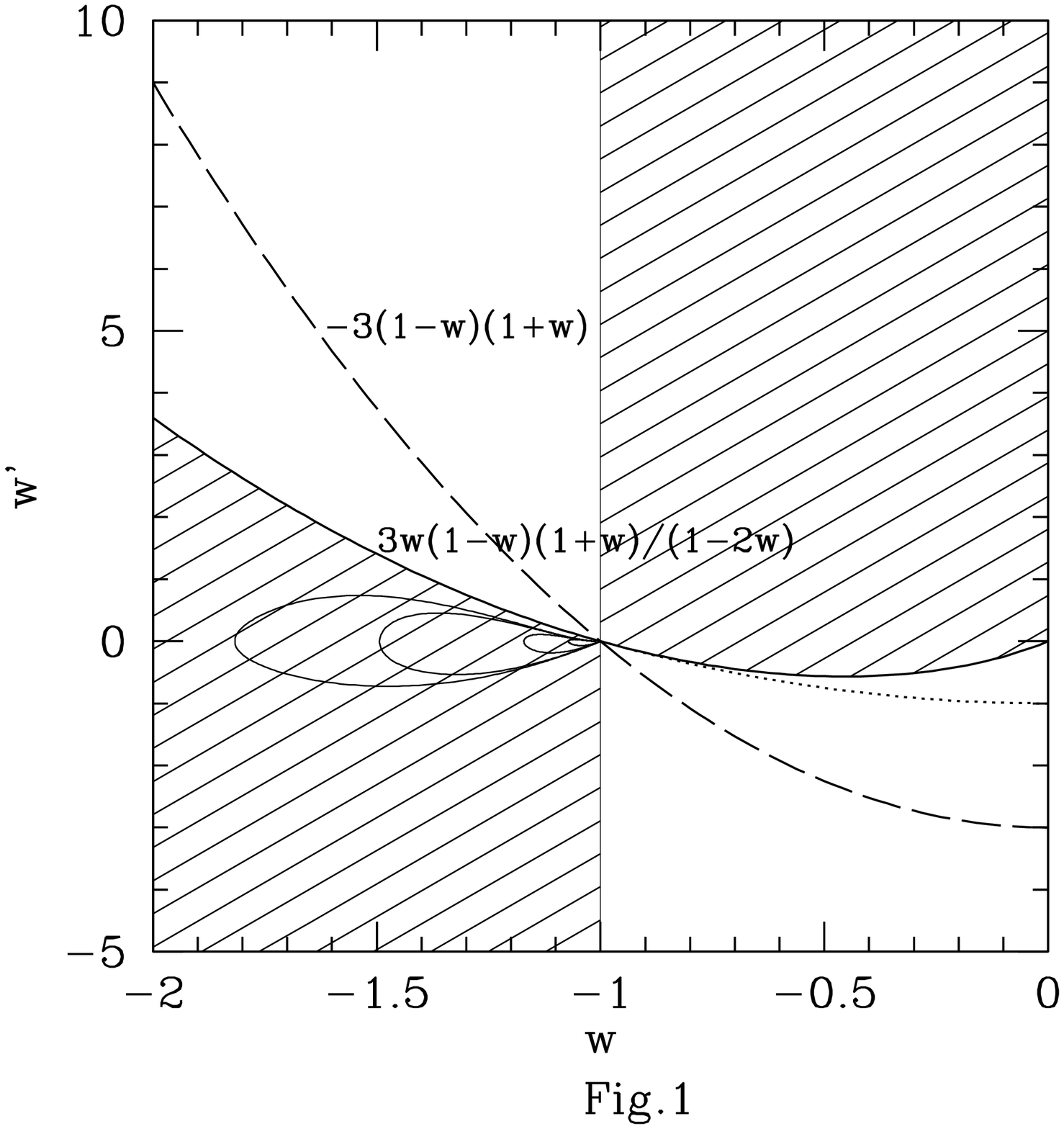}
\caption{Bounds on $w'$ as a function of $w$. For $w>-1$, curves are lower 
bounds: Solid curve is our lower bound while dotted curve is from Ref.\cite{sch}; 
while for $w<-1$ curves are upper bounds. 
Dashed curve is a generic lower/upper bound Eq.(\ref{quint}) and Eq. (\ref{ph:upper}). 
The shaded region is the bound for tracker fields, Eq. (\ref{q-track2}) and Eq. (\ref{ph-trac}). 
The loops in the shaded region 
are the trajectories for $V=V_0\log(\kappa\phi)$. } 
\label{fig:fig1}
\end{figure}

\begin{figure}
\includegraphics[width=\hsize]{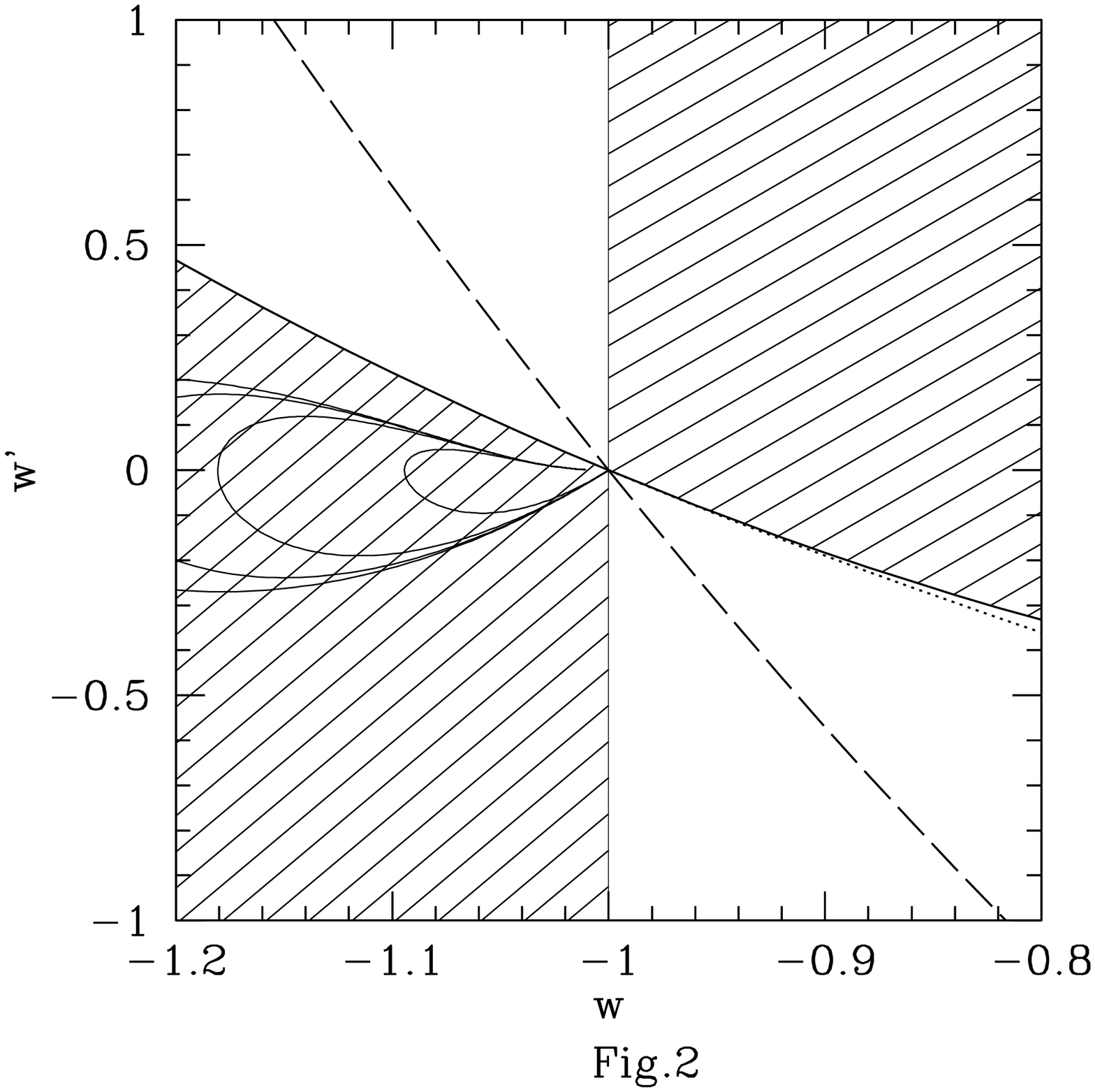}
\caption{Same as Fig. 1 but for narrower range of $w$. The loops on the shaded region 
are the trajectories for $V=V_0\sqrt{\kappa\phi}$.}
\label{fig:fig2}
\end{figure}

\section{Limits of Phantom}

We extend the range of $w$ to $w<-1$. 
Phantom is scalar field dark energy with $w<-1$ \cite{phantom}. 
Although there are several models of phantom, we consider a simple 
scalar field model with wrong sign of the kinetic term (ghost) \cite{phantom}. 
We understand that there are several serious obstacle (eg. rapid gravitational 
decay of vacuum) for such a ghost 
field to be cosmologically relevant field \cite{cht}. 
Our intention here is more observation-oriented and 
to provide observables related with the dynamics of the field 
which behaves like phantom. After giving the generic lower bound for phantom dark energy, 
we derive an upper bound on $w'$ for tracker phantom models.

\subsection{Generic Bound}

In this model, the energy density and the pressure of dark energy is given by 
$\rho=-\dot\p^2/2+V, p=-\dot\p^2/2-V$, respectively. We consider a non-negative $V(\phi)$.
The equation of motion is given by
\beq
\ddot\p+3H\dot\p-V_{,\phi}=0.
\eeq
Therefore, the scalar field rolls down the inverted potential $-V$ 
(or rolls up the potential $V$). 

Similar to quintessence, the phantom equation of motion can be rewritten as 
\beqa
\pm {V_{,\phi}\over V}=\sqrt{{-3\kappa^2(1+w)\over \Omega_{\phi}}}
\left(1+{x'\over 6}\right),
\eeqa
where the plus sign for $\dot\phi>0(V_{,\phi}>0)$ and the minus  sign for the 
opposite and $x$ is defined by
\beqa
x=\ln\left(-\frac{1+w}{1-w}\right).\nonumber
\eeqa
Since the left hand side is positive, we have a lower bound on $x'$, $x'>-6$, 
which in turn gives an upper bound of $w'$
\beq
w'<-3(1-w)(1+w).
\label{ph:upper}
\eeq

\subsection{Tracker Phantom}

The tracker equation for phantom field is given by
\beqa
\Gamma -1={3(w_B-w)(1-\Omega_{\phi})\over (1+w)(6+x')}-{(1-w)x'\over 2(1+w)(6+x')}
-{2x''\over (1+w)(6+x')^2}.
\label{eq:p-trac}
\eeqa
Therefore, for the tracker solution for which $w$ is nearly constant and 
$\Omega_{\phi}\rightarrow 0$, $w$ is given by 
\beq
w={w_B-2(\Gamma -1)\over 2(\Gamma -1) +1}.
\label{eosp}
\eeq
Thus $\Gamma<1/2$ is required for tracking phantom $w<-1$. 

Another issue to be addressed is the stability of the tracker solution 
against perturbation. In the case of quintessence, positive effective 
mass squared $V_{,\phi\phi}>0$ is required for the stability. 
For phantom the opposite condition 
$V_{,\phi\phi}<0$ is required since phantom rolls down the inverted 
potential. For constant $w$, to which a tracker solution corresponds, 
it can be shown that 
\beq
V_{,\phi\phi}=-{9\over 4}H^2(1-w)\left((1+\Omega_{\p})w+2\right).
\eeq
Thus, $V_{,\phi\phi}<0$ implies $-2<w<-1$ and in terms of $\Gamma$ 
the condition is  $\Gamma<0$. 
 
As tracker phantom models, we consider a solution in which $w$ is initially 
nearly constant and then 
it evolves toward $-1$. Such solutions are obtained for convex $V$, 
$V_{,\phi\phi}<0$ (a simple example is $V\propto \sqrt{\phi}$). 
This is because if $V_{,\phi\phi}<0$, then $|V_{,\phi}|$ decreases 
as $\phi$ climbs up the potential and $|V_{,\phi}|$ becomes negligible but 
$H$ increases as phantom becomes dominated and the 
dynamics of $\phi$ is eventually dominated by the Hubble friction 
($\ddot\phi\simeq -3H\dot\phi$) and $\phi$ ceases to move. 

Since $w'\geq 0$ for these tracker solutions, $x'\leq 0$. As phantom is 
attracted toward $w=-1$, $x'$ stops decreasing and then increases back to 
a value near zero, being 
similar to tracker quintessence \cite{sch}. The minimum value of $x'$, $x_m'$, 
gives an {\it upper bound} on $w'$ through Eq.(\ref{x'}). 

To find $x_m'$, we put $x''=0$ in Eq.(\ref{eq:p-trac}) and find that
\beqa
x_m'=-6{w(1-\Omega_{\phi})+2(1+w)(\Gamma -1)\over 
(1-w)+2(1+w)(\Gamma -1)}.
\eeqa
Since $x_m$ is an increasing function of $w(<-1)$, a lower bound is given 
by $w$ of the tracker solution Eq.(\ref{eosp}): 
\beq
x_m'>\frac{6w\Omega_{\p}}{1-2w}>\frac{6w}{1-2w}.
\eeq
{}From Eq.(\ref{x'}), in terms of $w'$, we obtain an upper bound on $w'$:
\beq
w'<\frac{3w}{1-2w}(1-w)(1+w).
\label{ph-trac}
\eeq
The bound is shown in Fig. 1 and Fig. 2. The trajectories 
of $(w,w')$ for a logarithmic potential $V=V_0\log(\kappa\p)$ (Fig. 1) and for 
$V=V_0\sqrt{\kappa\phi}$ (Fig. 2) are also shown for several initial conditions. 

\section{Limits of Tracker K-essence}

We finally give bounds on $w'$ for k-essence with $w>-1$. 
K-essence is a scalar field model of dark energy which has non-canonical 
kinetic term 
\cite{k-inf,coy,k-essence}. 
The pressure of the scalar field $\phi$, $p$, is given by the 
Lagrangian density $p(\p,X)$ itself where 
$X=-\partial_{\mu}\phi\partial^{\mu}\phi/2$ and is equal to $\dot\phi^2/2$ 
for the Friedmann model. The energy density $\rho$ is given by 
$\rho=2X\partial p/\partial X-p\equiv 2Xp_X-p$.

The equation of motion of the scalar field is given by 
\beqa
\ddot{\p}\left(p_X+
  \dot\p^{2}p_{XX}\right)+3Hp_X\dot\p+
p_{X\p}\dot\p^{2}-p_{\p} = 0,
\label{eomk}
\eeqa
where $p_{\p}=\partial p/\partial\p$, for example.
For the factorized form of 
\beq
p(\p,X)=V(\p)W(X),
\label{fact}
\eeq 
we can express the equation of motion 
of $\p$ in alternative form similar to quintessence \cite{ktrac}:
\beqa
\mp{V_{,\phi}\over V^{3/2}}&=& {\kappa \over 2}
\sqrt{{(1+w)W_X\over 3\Omega_{\p}} }
\left(6+   A x'\right),\label{eom}\\
A &=& {(XW_X-W)(2XW_{XX}+W_X)\over XW_X^2-WW_X-XWW_{XX}}={1-w\over
    c_s^2-w},
\label{eq:eomk}
\eeqa
where the minus (plus) sign corresponds to $\dot\p
>0(<0)$, respectively. $c_s^2$ is the speed of sound of k-essence defined by 
\cite{gm}
\beq
c_s^2={\delta p\over \delta\rho}={p_X\over p_X+2Xp_{XX}}
={W_X\over W_X+2XW_{XX}}.
\eeq

\subsection{Tracker K-Essence}
Similar to quintessence, we define a dimensionless function $\Gamma$ 
by  $\Gamma =VV_{,\phi\phi}/V_{,\phi}^2$. Taking 
the time derivative of Eq.(\ref{eom}), we obtain \cite{ktrac}
\beqa
&&\Gamma-{3\over 2}=-{1\over (1+w)(6+A x')}\left[
3(w-w_B)(1-\Omega_{\p}) +{(1-w)^2\over 
2(c_s^2-w)} x'\right. \nonumber\\
&&+\left.{2(1-w)(c_s^2-w) x'' + 2\left(\dot w(1-c_s^2)-
(c_s^2)^{\cdot}(1-w)\right) x'/H
\over (6+A x')(c_s^2-w)^2}
\right].\label{tracker}
\eeqa
 Eq.(\ref{tracker}) might be called 
the k-essential counterpart of the tracker equation.
Therefore for the tracker solution (assuming $\Gamma \simeq$ const. 
and $\Omega_{\p}\ll 1$) 
we can write $w$ in terms of $\Gamma$:
\beq
w={w_B-2(\Gamma -3/2)\over 2(\Gamma -3/2)+1}\simeq {\rm const}.\label{eos}
\eeq
For tracker k-essence models with $w>-1$, 
$w$ is nearly constant, so $X$ is also constant 
since $w$ only depends on $X$ for the factorized $p(\p,X)$, Eq.(\ref{fact}). 
Then the evolution of energy density depends only on $V(\p)$, so  
$\dot\p>0(<0)$ corresponds to $V_{,\phi}<0(>0)$ and  the left hand side 
of Eq.(\ref{eq:eomk}) is positive. This implies $6+Ax'>0$, which is written 
in terms of $w'$ as
\beq
w'>-3(c_s^2-w)(1+w).
\eeq 
This bound is similar to Eq.(\ref{quint}). However, in deriving it, we 
restrict ourselves to tracker k-essence models. 

Similar to tracker quintessence, we can sharpen the bound by considering 
the dynamics more carefully. 
Since $w'\leq 0$ for tracker fields, $x'\leq 0$. However, since $w$ 
asymptotically approaches toward $-1$, $x'$ eventually stops decreasing and 
then increases toward zero. The minimum of $x'$, $x_m'$, gives the minimum of 
$w'$ via Eq.(\ref{x'}). The analysis is the same as that of tracker 
quintessence and we only give the final result:
\beqa
w'>{3w\over 1-2w}(c_s^2-w)(1+w).
\eeqa
This is the k-essential counterpart of the lower bound on $w'$. 
If we impose the upper bound on the sound speed as $c_s^2\leq 1$, then
the above bound is reduced to that of tracker quintessence:
\beqa
w'>{3w\over 1-2w}(1-w)(1+w).
\eeqa
The sound speed of dark energy \cite{csn,hu,cs} is currently difficult to 
measure (see for example, \cite{hannestad}). Therefore, it seems difficult to 
distinguish quintessence from k-essence from the measurements of $w$ and $w'$. 

\section{Summary}

In this paper, we have extended and generalized bounds on dark energy 
models in the $w-w'$ plane. First, we have slightly improved the lower bound 
for tracker quintessence. Second, we have derived both 
an upper bound for  general phantom models and an upper bound for tracker phantom. 
Finally, we have obtained two lower bounds for k-essence. 
While the required observational accuracy of $w'$ is similar for 
quintessence and k-essence, $\sigma(w')\siml (1+w)$, the windows of $w'$ 
for phantom may be similar to that of quintessence, $\sigma(w')\siml |1+w|$. 
Although the fate of the universe with $w<-1$ would be disastrous (the 
future big rip singularity and the disintegration of bound objects) 
\cite{ckw,cts}, be aware of the possibility. 

\begin{acknowledgments}
This work was supported in part by a Grant-in-Aid for Scientific 
Research (No.15740152 and No.17204018) from the Japan Society for the 
Promotion of Science and in part by Nihon University. 
The author would like to thank C.-W. Chen, J.-A. Gu and P. Chen \cite{cgc} for 
pointing out errors in the previous version of the paper. 
\end{acknowledgments}


\end{document}